\documentclass[a4paper,11pt,preprintnumbers]{article}

\usepackage{pos}
\usepackage{graphicx}
\usepackage{amsmath}
\usepackage{amsfonts}
\usepackage{bbold}
\usepackage{slashed}

\newcommand{\ignore}[1]{}
\newcommand\SU{\mathrm{SU}}

\newcommand{\cL}{\ensuremath{\mathcal L} }

\newcommand{\cS}{\ensuremath{\mathcal S} }
\newcommand{\cD}{\ensuremath{\mathcal D} }
\newcommand{\Tr}[1]{\ensuremath{\mbox{Tr}\left[ #1 \right]}}

\newcommand{\matrixel}[3]{\left< #1 \vphantom{#2#3} \right|
	#2 \left| #3 \vphantom{#1#2} \right>} 

\title{Hidden Conformal Symmetry from Eight Flavors}

\author*{James Ingoldby}
\onbehalf{on behalf of the Lattice Strong Dynamics Collaboration}

\affiliation{Abdus Salam International Centre for Theoretical Physics, Strada Costiera 11, 34151, Trieste, Italy}
\affiliation{Institute for Particle Physics Phenomenology, Durham University, Durham DH1 3LE, United Kingdom}

\emailAdd{james.a.ingoldby@durham.ac.uk}

\abstract{This proceedings paper extends the scope of our conference talk, where we presented a comprehensive analysis of newly expanded and refined lattice data concerning the $\SU(3)$ gauge theory with $N_f=8$ light Dirac fermions — a theory positioned near the conformal window boundary. The analysis presented here makes use of a dilaton effective field theory and we delve deeper into the intricacies of the dilaton potential. We aim to clarify the connection between parameters appearing the potential and properties of the underlying gauge theory.}

\FullConference{The 40th International Symposium on Lattice Field Theory (Lattice 2023)\\
July 31st - August 4th, 2023\\
Fermi National Accelerator Laboratory\\}

\begin{document}
\maketitle

\section{Introduction}

The $\SU(N_c)$ gauge theories with $N_f$ flavors of Dirac fermions constitute a varied family of theories, including many members that markedly differ from QCD. For a given $N_c$, there exists a critical number of fermions $N_{fc}$ at which the gauge theory makes a transition from a confining phase, to another phase characterized by conformal behavior in the infrared. Distinguishing between these behaviors requires non-perturbative calculations with strong gauge coupling.

Non-perturbative lattice studies of the $\SU(3)$ gauge theory with $N_f = 8$ have suggested that it is close to the transition point for conformal behavior~\cite{Appelquist:2007hu,Deuzeman:2008sc,Fodor:2009wk,Hasenfratz:2014rna,Fodor:2015baa,Aoki:2016wnc,Appelquist:2016viq,Appelquist:2018yqe,Kotov:2021mgp,Hasenfratz:2022zsa,Hasenfratz:2022qan,LSD:2023uzj}, so that $N_{fc}$ is close to 8. Lattice studies also indicate the presence of an unexpectedly light scalar state in the spectrum~\footnote{Lighter than the $\rho$ resonance and lighter than its analog in $N_f=2$ QCD when lattice data taken using comparable fermion masses are considered \cite{LatticeStrongDynamics:2023bqp,Rodas:2023twk}.}. The lightness of the scalar could be explained if it were an approximate dilaton, the Goldstone boson that arises when scale invariance is spontaneously broken. Potentially, we might expect gauge theories close to the conformal transition point to possess an approximate scale invariance (above their confinement scales), which would be needed to furnish a dilaton.

In these proceedings, we extend the study of Ref.~\cite{LSD:2023uzj}, where we analyze recent lattice data for the $N_f = 8$ theory using a dilaton Effective Field Theory (dEFT). In particular, we comment on issues regarding the form of a potential appearing in the dEFT Lagrangian and its connection to features of the underlying $N_f=8$ gauge theory. Dilaton EFT descriptions of confining gauge theories close to the conformal transition have been developed in Refs.~\cite{Golterman:2016lsd,Golterman:2016hlz,Fodor:2017nlp,Golterman:2018mfm,Fodor:2020niv,Golterman:2020utm,Golterman:2021ohm,Freeman:2023ket,Appelquist:2022mjb,Zwicky:2023krx} and earlier lattice datasets of the $\SU(3)$, $N_f=8$ theory have been fitted well by dEFT \cite{Appelquist:2017wcg,Appelquist:2019lgk,Fodor:2019vmw,Golterman:2020tdq}. By using dEFT, we assume that $N_f = 8$ is confining rather than conformal in the infrared. An alternative analysis of the same lattice dataset that assumes conformal behavior was presented in \cite{LSD:2023uzj}. 

Dilaton EFTs have been used in many contexts besides the analysis of lattice data. See Ref.~\cite{Coleman:1985rnk} for an early introduction. Indeed, the Higgs boson itself can be thought of as an approximate dilaton. Fundamentally, the Higgs boson may be a dilaton that originates from a new strongly coupled conformal sector. See, for example Ref.~\cite{Goldberger:2007zk}. Alternatively, it may form within the strong sector as an admixture between the dilaton and another Goldstone boson and be described using dEFT~\cite{Appelquist:2020bqj}.

In the following two subsections, we briefly review our formulation of dEFT and the set of lattice data which we will be fitting using our EFT. In Section.~\ref{sec:fit}, we describe our fits to data, presenting new results for the favored range of a particular parameter that enters the dilaton potential. Finally in Section.~\ref{sec:interp-delta}, we explain our derivation of the dilaton potential and justify its form.

\subsection{Dilaton EFT}

Within dEFT, the dilaton field $\chi$ acquires a nonzero vacuum expectation value (VEV) $\langle\chi\rangle = F_d$, spontaneously breaking scale invariance. The EFT also contains a multiplet of pNGBs arising as pseudo Goldstone bosons of the approximate chiral symmetry $\SU(N_f)_L\times\SU(N_f)_R$, which gets spontaneously broken to its subgroup $\SU(N_f)$. Under chiral symmetry, the pNGB multiplet $\Sigma$ transforms as a bifundamental $\Sigma\rightarrow L\Sigma R^\dagger$ for $L,\,R\in\SU(N_f)_{L,\,R}$. The spontaneous breaking of chiral symmetry is achieved by giving the pNGB field the VEV $\langle\Sigma\rangle=\mathbb{1}$, with fluctuations satisfying the constraint $\Sigma^\dagger\Sigma=\mathbb{1}$.

The dEFT Lagrangian is reviewed in Ref.~\cite{Appelquist:2022mjb}. At leading order in derivatives and explicit symmetry breaking interactions, it is given by
\begin{align}
	\cL = \frac{1}{2}\partial_{\mu}\chi\partial^{\mu}\chi \, + \frac{f_{\pi}^2}{4}\left(\frac{\chi}{f_d}\right)^2 \, \Tr{\partial_{\mu}\Sigma(\partial^{\mu}\Sigma)^{\dagger}} + \frac{m_{\pi}^2 f_{\pi}^2}{4} \left(\frac{\chi}{f_d}\right)^y \, \Tr{\Sigma + \Sigma^{\dagger}} \, - \, V(\chi) \, .
	\label{eq:L}
\end{align}
The first two terms in the Lagrangian preserve scale and chiral symmetries exactly, serving as kinetic terms with two derivatives. The third term breaks scale and chiral symmetries explicitly and provides mass for the pNGBs. The origin of this breaking is the fermion mass $m$ in the underlying gauge theory, so we identify $m^2_\pi=2B_\pi m$. The parameter $y$ should be identified with the scaling dimension of the fermion bilinear in the underlying gauge theory \cite{Leung:1989hw}. At the bottom of the conformal window, it has been argued that $y\sim 2$ - see Ref.~\cite{Zwicky:2023bzk} and references therein.

We include a potential for the dilaton, of the form
\begin{align}
	V_{\Delta}(\chi) = \frac{m_d^2\chi^4}{4(4-\Delta)f_d^2}\left[1-\frac{4}{\Delta}\left(\frac{\chi}{f_d}\right)^{\Delta-4} \right] \, .
	\label{eq:vdelta}
\end{align}
Potentials of this type have been discussed in Refs.~\cite{Rattazzi:2000hs,Goldberger:2007zk,Chacko:2012sy}. The potential includes a scale invariant term $\propto \chi^4$, and a term which breaks scale invariance $\propto\chi^\Delta$. The latter descends from interactions in the underlying gauge theory that explicitly break scale but not chiral invariance and can be motivated using a spurion analysis \cite{Appelquist:2022mjb}.

In these proceedings, we take $\Delta$ to be a free parameter to be determined from fits to lattice data. In the limit $\Delta\rightarrow4$, the potential in Eq.~(\ref{eq:vdelta}) takes the logarithmic form
\begin{align}
	V_4(\chi) = \frac{m_d^2}{16f^2_d}\chi^4\left[4\log\left(\frac{\chi}{f_d}\right)-1\right]\,.
	\label{eq:vlog}
\end{align}
For this choice of potential, our dEFT coincides with that developed in Ref.~\cite{Golterman:2016lsd}.

By expanding fields around the classical vacuum, we can compute the leading-order physical dilaton and pNGB masses, $M^2_d$ and $M^2_\pi$, as implicit functions of the symmetry-breaking quantity $m^2_\pi=2B_\pi m$. Similarly, the leading-order pNGB decay constant $F_\pi$ can be computed using the axial vector Noether current.

Additionally, we calculate a scalar decay constant $F_{S}$, defined through the matrix element
\begin{equation}
	\matrixel{0}{J_{S}(x)}{\chi(p)} \equiv F_{S} M_d^{2}e^{-ip\cdot x}\,,
\end{equation}
where the scalar current $J_S=m\sum_{i=1}^{N_f}\bar{\psi}_i\psi_i$ in the underlying gauge theory. In the EFT, the decay constant $F_S$ is given by \cite{LSD:2023uzj}
\begin{equation}
	|F_S| = \frac{yN_fM^2_\pi F_\pi}{2M^2_d}\frac{f_\pi}{f_d}.
	\label{eq:dilfspred}
\end{equation}
This equation, which has also been derived using current algebra techniques \cite{LatKmi:2015non}, plays a crucial role in our analysis.

Furthermore, in our study, we incorporate scattering parameters. In dEFT, the scattering length $a^{(2)}_{0}$ for pNGBs in the s--wave, maximal isospin channel has been found to be \cite{LatticeStrongDynamicsLSD:2021gmp}:
\begin{align}
	M_\pi a^{(2)}_{0} = -\frac{M^2_\pi}{16\pi F^2_\pi}\left(1-(y-2)^2\frac{f^2_\pi}{f^2_d}\frac{M^2_\pi}{M^2_d}\right)\,.\label{eq:dilatonslength}
\end{align}
The first term in this expression corresponds to the expression for scattering in chiral perturbation theory in this channel \cite{Weinberg:1966kf}, even in the absence of a dilaton. The second term, proportional to $M^2_\pi/M^2_d$, represents the contribution from the dilaton, which is suppressed for $y\sim2$.

\subsection{Lattice Data for the $\SU(3)$ Gauge Theory with $N_f=8$ Flavors}

In our analysis of the dilaton effective field theory (dEFT), we rely on lattice data obtained from the comprehensive study presented in the recent Lattice Strong Dynamics collaboration paper \cite{LatticeStrongDynamics:2023bqp}. This dataset encompasses four crucial quantities, namely the masses for the pseudoscalar Nambu-Goldstone boson ($aM_\pi$) and the dilaton ($aM_d$), along with the decay constants ($aF_\pi$ and $aF_S$). The lattice spacing, denoted by $a$, serves as a convenient unit of measurement for dimensionful quantities obtained through lattice calculations. The data has been extrapolated to the infinite volume limit and is conveniently summarized in Table IX of \cite{LatticeStrongDynamics:2023bqp}, with additional details on lattice action and ensembles available in the same source.
The dataset includes measurements for these quantities across five distinct values of the fermion mass expressed in lattice units ($am$). 

Additionally, we incorporate lattice data for the s-wave scattering phase shift in the $I=2$ isospin channel, obtained from Ref.~\cite{LatticeStrongDynamicsLSD:2021gmp}. The observable we utilize is $M_\pi/k\cot\delta$, which, in the low momentum limit ($k\ll M_\pi$), coincides with the scattering length $M_\pi a^{(2)}_0$. Although the dataset in \cite{LatticeStrongDynamicsLSD:2021gmp} is presented in terms of $M_\pi/k\cot\delta$, we interpret these measurements as data for the scattering length in our EFT fits. In total, we have data for 5 observables at 5 different values of the fermion mass, yielding 25 data points total.

To compare EFT fits with varying numbers of model parameters, we employ the Akaike Information Criterion (AIC) \cite{Akaike:1974AIC}. The AIC is defined as $\text{AIC} = \chi^2_\text{min} + 2k$, where $\chi^2_\text{min}$ is the minimum chi-squared value and $k$ represents the number of model parameters. Our analysis reveals that the statistical covariances between different observables are negligible, which simplifies the calculation of the chi-squared function.

In a Bayesian framework, the relative probability of two models being correct can be determined by comparing their AIC values \cite{Jay:2020jkz}. Specifically, the probability ratio $p_1/p_2$ is given by the formula
\begin{align}
	p_1/p_2 = \exp\left(\left(\text{AIC}_2-\text{AIC}_1\right)/2\right)\,.
	\label{eq:relp}
\end{align}
This comparison provides valuable insights into the likelihood of different models being an accurate representation of the underlying physics.

\section{Fits to Lattice Data}
\label{sec:fit}

We show the results of two global fits of leading-order dEFT to our lattice data in Table~\ref{Tab:deft}. We first perform a fit using the form in Eq.~(\ref{eq:vdelta}) for the dilaton potential. The results of this six-parameter fit are shown in the middle column. These results were reported in Ref.~\cite{LSD:2023uzj} and are consistent with earlier determinations \cite{Appelquist:2017wcg,Appelquist:2019lgk,Fodor:2019vmw}. After minimizing the chi-squared function with respect to these 6 parameters, we obtain a chi-squared minimum per degree of freedom just over 1, indicating an acceptable quality of fit. 

For comparison, we perform a fit using the logarithmic potential shown in Eq.~(\ref{eq:vlog}), which is the form Eq.~(\ref{eq:vdelta}) takes in the $\Delta\rightarrow4$ limit, when our EFT coincides with that of \cite{Golterman:2016lsd}. Results for this five-parameter fit are shown in the last column of Table~\ref{Tab:deft}. Again, the chi-squared minimum per degree of freedom lies just above 1, indicating a reasonable fit.

To make a more quantitative comparison between models, we use the AIC. By plugging values from Table.~\ref{Tab:deft} into Eq.~(\ref{eq:relp}), we find that the relative probability for the two models is $p_4/p_\Delta = 0.18$, which mildly disfavors the $\Delta\rightarrow4$ version of the EFT as a model describing this dataset.

\begin{table}[t]
	\centering
	\vspace{12pt}
	\renewcommand\arraystretch{1.2}
	\begin{tabular}{| c || c | c |}
		\hline
		Parameter & $V_\Delta$ Fit & $V_4$ Fit\\
		\hline\hline
		$y$ & 2.091(32) & 2.087(32)\\
		$B_\pi$ & 2.45(13) & 2.16(19)\\
		$f^2_\pi$ & $6.1\,(3.2)\times10^{-5}$ & $3.0\,(2.2)\times10^{-6}$\\
		$f_\pi^2/f_d^2$ & 0.1023(35) & 0.1011(35) \\
		$m_d^2 / f_d^2$ & 1.94(65) &  0.499(54)\\
		\hline
		$\Delta$ & 3.06(41) & ---\\
		\hline\hline
		$\chi^2/\text{dof}$ & 21.3 / 19 & 26.7 / 20\\
		AIC & 33.3 & 36.7 \\
		\hline
	\end{tabular}
	\caption{Central values and uncertainties for the leading-order $V_\Delta$ (six-parameter) and $V_4$ (five-parameter) fits to the dEFT. Lattice data for $M_{\pi}$, $M_d$, $F_\pi$, $F_S$ and the $I=2$ scattering length has been incorporated into this fit, for 5 different vales of the underlying fermion mass $m$, corresponding to 25 data points. All dimensionful quantities are presented in units of the lattice spacing.}
	\label{Tab:deft}
\end{table}

\begin{figure}[t]
	\begin{center}
		\includegraphics[width=0.6\columnwidth]{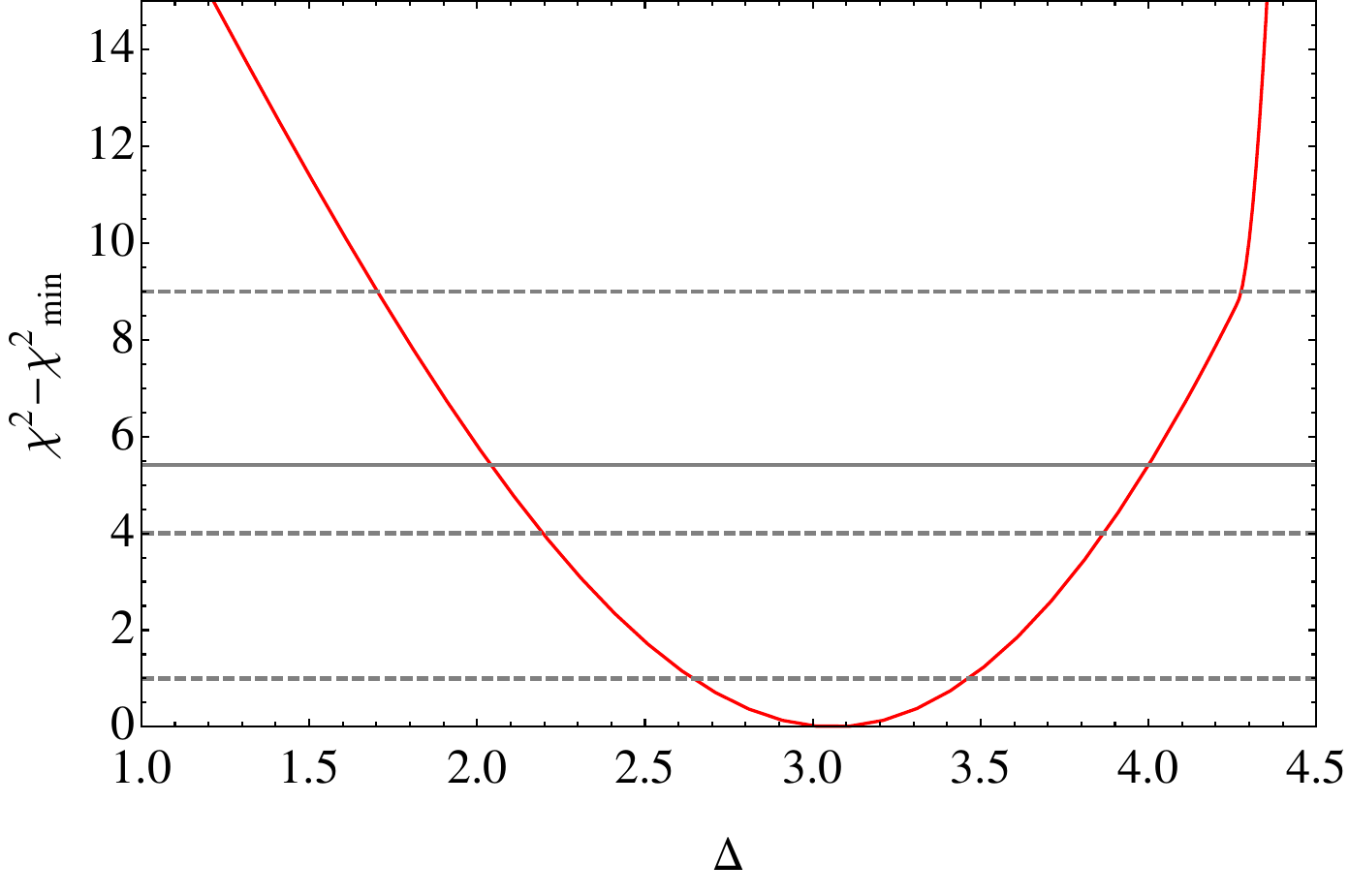}
	\end{center}
	\caption{The $\chi^2$ function plotted against $\Delta$ after minimizing with respect to the remaining 5 fit parameters.}
	\label{Fig:chidelta}
\end{figure}

To determine the favored range for $\Delta$, we plot the chi-squared function for the six parameter fit against $\Delta$ (after having minimized the chi-squared with respect to the remaining 5 EFT parameters) in Fig.~\ref{Fig:chidelta}. The three gray dashed lines correspond to the contours $\Delta\chi^2=1,\,4,\,9$ and indicate the extent of the 1, 2 and 3$\sigma$ ranges for $\Delta$. For example, the red curve crosses the $\Delta\chi^2=1$ contour at $\Delta=2.6$ and 3.5 giving a 1$\sigma$ range for $\Delta$ between 2.6 and 3.5. The solid line crosses the red at $\Delta\rightarrow4$, showing that this limit sits between the 2 and 3$\sigma$ ranges for $\Delta$.

\section{Interpretation of $\Delta$}
\label{sec:interp-delta}

We begin this section by summarizing the derivation of the dilaton potential in Eq.~(\ref{eq:vdelta}), clarifying the role of the parameter $\Delta$ in our construction. We then go on to describe how $\Delta$ relates to quantities in the underlying gauge theory and explain why values of $\Delta$ below 4 emerge naturally given this UV completion, in contrast to the claim made in Ref.~\cite{Golterman:2020utm}.

We derive the potential using a spurion analysis, following the logic outlined in Refs.~\cite{Appelquist:2019lgk,Appelquist:2022mjb}. We start with a dEFT that is scale invariant. This dEFT then has a moduli space of degenerate vacua in which scale invariance is spontaneously broken by the dilaton field acquiring a VEV. To realistically describe nearly conformal lattice gauge theories at low energies, there needs to be some (weak) explicit breaking of scale invariance built into the dEFT, to give the dilaton a mass. This is implemented by introducing a spurion field $\cS(x)$, which transforms under scale transformations $x\rightarrow e^\rho x$ with the rule 
\begin{align}
	\cS(x)\rightarrow e^{(4-\Delta)\rho}\cS(e^\rho x)\,.
	\label{eq:sptrans}
\end{align}
At this point, $\Delta$ appears inside the scaling dimension of the spurion field - its quantum number for scale invariance.

The dilaton potential is then built out of scale invariant combinations of the spurion and dilaton fields. In addition, we require that the potential is analytic in the spurion. It takes the form
\begin{align}
	V(\chi) = \chi^4\sum_{n=0}a_n\left(\frac{\cS}{\Lambda^2}\left(\frac{\chi}{f_d}\right)^{\Delta-4}\right)^n\,,
	\label{eq:vfull}
\end{align}
where the $a_n$ are dimensionless constants, $\Lambda$ is the dEFT cutoff (given roughly by the confinement scale of the gauge theory), and we have normalized $\cS$ to give it dimensions of mass squared.

Scale invariance is then explicitly broken by demoting the spurion from a field that transforms with rule (\ref{eq:sptrans}), to a constant $m^2_d$ that doesn't transform. Provided that $m^2_d\ll\Lambda^2$, we can truncate the sum in Eq.~(\ref{eq:vfull}). At leading order in the low energy expansion, the potential for the dilaton takes the form shown in Eq.~(\ref{eq:vdelta}).

We stress that to derive the dilaton potential in dEFT, we have invoked only the low energy degrees of freedom, the symmetries that act on them and the spurions that break these symmetries. We have not made any direct use of information about the underlying theory above the confinement scale (e.g the gauge group, fermion count or beta function for the gauge coupling).

In principle, further constraints on $\Delta$ however come from the UV. Spurions must correspond to couplings of symmetry breaking terms present in a generating functional renormalized at the scale where the EFT is matched to the UV theory. Since we know that the UV theory is the $\SU(3)$ $N_f=8$ gauge theory, we have schematically
\begin{align}
	\exp\{-Z[\cS]\} = \int\cD\bar{\psi}\cD\psi\cD G_\mu\exp\{-S_\text{g.t}[\cS]\} = \int\cD\Sigma\cD\chi\exp\{-S_\text{dEFT}[\cS]\}\,,
	\label{eq:gfunc}
\end{align}
where $S_\text{g.t}$ is the action of the $N_f=8$ gauge theory. The spurion $\cS$ has been introduced into $S_\text{g.t}[\cS]$ in such a way so as to make the generating functional scale invariant, if $\cS$ is given the transformation law shown in Eq.~(\ref{eq:sptrans}). This dependence is analytic to ensure that correlation functions can be extracted from the generating functional by taking functional derivatives with respect to $\cS$.

At the confinement scale where matching with the EFT happens, the gauge theory is strongly coupled. Furthermore in a gauge theory near the conformal transition, the coupling is expected to remain strong over an extended interval of scales above the confinement scale, with small beta function. There is then the possibility that operator anomalous dimensions become large. This is seen already in the lattice data, since $y \approx 2$ can be identified with the $\bar{\psi}\psi$ scaling dimension, which is very different from its engineering dimension.

If under scale transformations, the only change to the gauge theory generating functional were to come from the running of the dimensionless gauge coupling $g(\mu)\rightarrow g(\mu+\delta\mu)$, then in effect the theory is scale invariant with a near marginal deformation. However this cannot be assumed at strong coupling, since the large anomalous dimensions that arise can result in new gauge theory operators becoming relevant. These new relevant operators should then be included in the renormalized action (that enters the generating functional), even though they were not present originally in the bare action of the gauge theory.

The possibility that there may be new relevant operators appearing in the gauge theory just below the conformal window has been widely considered in the literature \cite{Gies:2005as,Kaplan:2009kr}. In particular, it has been suggested that the transition between the infrared conformal and confining phases of gauge theories may be caused by a four--fermi operator becoming relevant\footnote{Due to operator mixing under RG flows, an operator that becomes relevant at strong coupling could be an admixture of many operators, and the four--fermi would only be the largest component.}. In which case, this would be an example of a ``marginality crossing transition'' \cite{Gukov:2016tnp}. A lattice measurement of a four--fermi operator anomalous dimension in a conformal gauge theory is presented in Ref.~\cite{DelDebbio:2013uaa}.

In the marginality crossing case, a new relevant operator should be included in $S_\text{g.t}$ inside Eq.~(\ref{eq:gfunc}). This new operator will come with a new coupling, that should be promoted to a spurion to restore scale invariance, which we then identify with $\cS$. In this case, the EFT parameter $\Delta$ ought to be identified with the scaling dimension of the new relevant operator at the confinement scale, which in general will not be close to 4. In order to analyze the lattice data, while making as few assumptions as possible about the nature of the conformal transition, we take $\Delta$ to be a free parameter, to be determined from fits to lattice data.


\acknowledgments

We would like to thank Roman Zwicky and Maurizio Piai for useful discussions, as this work was being completed.

\bibliographystyle{JHEP}
\bibliography{nf8short}

\end{document}